\documentclass[]{pasj01}

\usepackage{threeparttable}
\usepackage[referable]{threeparttablex}
\usepackage{booktabs}
\usepackage{multirow}
\usepackage{color}
\usepackage{adjustbox}
\usepackage{lscape}
\usepackage{longtable}
\usepackage{comment}

\Received{$\langle$reception date$\rangle$}
\Accepted{$\langle$acception date$\rangle$}
\Published{$\langle$publication date$\rangle$}

\begin{document}

\title{
Study on the escape timescale of high-energy particles from supernova remnants through thermal X-ray properties
}

\author{Hiromasa Suzuki$^1$, Aya Bamba$^{1, 2}$, Ryo Yamazaki$^{3,4}$, Yutaka Ohira$^5$}

\altaffiltext{1}{Department of Physics, The University of Tokyo, 7-3-1 Hongo, Bunkyo-ku, Tokyo 113-0033, Japan}
\altaffiltext{2}{Research Center for the Early Universe, School of Science, The University of Tokyo, 7-3-1 Hongo, Bunkyo-ku, Tokyo 113-0033, Japan}
\altaffiltext{3}{Department of Physics and Mathematics, Aoyama Gakuin University, 5-10-1 Fuchinobe, Chuo-ku, Sagamihara, Kanagawa 252-5258, Japan }
\altaffiltext{4}{Institute of Laser Engineering, Osaka University, 2-6 Yamadaoka, Suita, Osaka 565-0871, Japan}
\altaffiltext{5}{Department of Earth and Planetary Science, The University of Tokyo, 7-3-1 Hongo, Bunkyo-ku, Tokyo 113-0033, Japan}

\email{suzuki@juno.phys.s.u-tokyo.ac.jp}


\maketitle

\begin{abstract}
In this decade, GeV/TeV gamma-ray observations of several supernova remnants (SNRs) have implied that accelerated particles are escaping from their acceleration sites. However, when and how they escape from the SNR vicinities are yet to be understood. Recent studies have suggested that the particle escape might {develop} with thermal plasma ages of the SNRs.
In this paper, we present a systematic study on time evolution of particle escape using thermal X-ray properties and gamma-ray spectra.

We used 38 SNRs which associate with GeV/TeV gamma-ray emissions.
We conducted spectral fittings on the gamma-ray spectra {using exponential cutoff power law and broken power law models to estimate the exponential cutoff or the break energies}, both of which are indicators of particle escape.
{The plots of the gamma-ray cutoff/break energies over the plasma ages show similar tendencies to those predicted by simple theories} of the particle escape under conditions in which a shock is interacting with thin interstellar medium or clouds.
{The particle escape timescale is estimated as $\sim$100 kyr from decreasing trends of the total energy of the confined protons with the plasma age.}
The large dispersions of the cutoff/break energies of the data may suggest an intrinsic variety of particle escape environments. This might be the cause of the complicated Galactic cosmic-ray spectral shape measured on Earth.

\end{abstract}

\section{Introduction}

Recent studies have revealed that charged particles are accelerated to the energies above TeV at young or middle-aged supernova remnant (SNR) shocks (e.g. Tycho: \cite{giordano12}; Cassiopeia A: \cite{ahnen17}; RX J1713.7-3946: \cite{hess18}). Even these seemingly most powerful accelerators in our Galaxy have not been found to provide particles of up to 10$^{15.5}$ eV, which might suggest very-high-energy particle escape from these SNRs \citep{ohira10}.
In the cases of middle-aged to old SNRs, most of their spectra show sharp cutoffs around several GeV (e.g. W44: \cite{ackermann13}; HB 21: \cite{ambrogi19}), indicating that higher-energy particles have already escaped from the SNRs \citep{ohira11a}. Also, most of the middle-aged to old gamma-ray emitting SNRs show the pion-decay spectral feature below 1 GeV, favoring the hadronic origins for the gamma-ray emissions.
And in some cases, GeV/TeV emissions from the adjacent clouds have been detected, suggesting the irradiation by escaped particles (W44: \cite{uchiyama10}; W28: \cite{cui18}).
However, when and how they escape from the SNR vicinities remain unknown and should be addressed observationally.

Interestingly, particle escape from SNRs have been suggested to be somehow related to {recombining plasmas (RPs)}, since most of the SNRs with RPs have the gamma-ray counterparts \citep{suzuki18}.
RPs are plasmas with abnormally low electron temperatures and thus in recombination-dominant states, which are opposite from ionizing plasmas (IPs) usually found in SNRs.
These RPs have been found in 16 middle-aged to old SNRs until now (e.g. IC 443: \cite{yamaguchi09}; W 49 B: \cite{ozawa09}; G359.1-0.5: \cite{ohnishi11}; W 28: \cite{sawada12}). The origin of the RPs are still under discussion.

Comparing the GeV gamma-ray and thermal X-ray data, \citet{suzuki18} suggested a possibility that particle escape and generation of RPs share the same origin, since the particle escape apparently {develops} with the RP age.
\citet{zeng19} have revealed that the GeV/TeV gamma-ray emitting SNRs show gradual particle escape with the SNR ages.
Thus, particle escape seemingly {develops} with an elapsed time from the supernova explosion, and the escape progress may be able to be tracked by the thermal plasma ages.

In this study, we conduct a systematic study on the gamma-ray emitting SNRs to derive the escape timescale from the observations, and make comparisons with theoretical predictions.
We suggest that the plasma age represents the SNR age and thus can be used to track the particle escape progress.
In Section \ref{sec-data}, we briefly describe the sample selection and show the analysis results on the GeV/TeV spectral fittings. In Section \ref{sec-sysanalysis}, we present a systematic study on the time evolution of particle escape. The results are discussed in Section \ref{sec-discussion}, and summarized in Section \ref{sec-summary}.
Throughout this paper, errors in the text, tables and figures represent a 1 $\sigma$ confidence level.

\section{Sample selection and gamma-ray spectral analysis}\label{sec-data}

\subsection{Sample and physical parameters}
We selected the gamma-ray emitting SNRs shown in \citet{acero16} or \citet{zeng19}.
Among them, we used objects with individually published gamma-ray spectra. We found 38 available objects, which are shown in Table \ref{tab-snrs}.
{We took the SNR diameters from Green's SNR catalogue \citep{green17} for the sake of uniformity.}
For the thermal plasma properties, we took electron number densities $n_{\rm e}$, and for those whose plasmas have deviations from collisional ionization equilibrium (CIE), we also extracted the ionization/recombination timescales $n_{\rm e}t$ from the literature and calculated the plasma ages ($t_{\rm p}$) by dividing $n_{\rm e}t$ with $n_{\rm e}$. The $n_{\rm e}$ is estimated from thermal X-ray emission measures and assumptions of the X-ray emitting volumes. The $n_{\rm e}t$ is estimated in the X-ray spectroscopies. We basically used the $n_{\rm e}t$ for the whole SNR. For the IPs or RPs detected from several regions, we used the average values.
Note that the origins of the RPs are still unclear, but since a correlation between the progress of particle escape and the RP ages has been suggested \citep{suzuki18, katsuragawa19, zhang19}, here we assumed that the RP ages had a positive correlation with the SNR ages.
From an SNR diameter, post-shock density $n_{\rm e}$ (assumed to be uniform), and an assumed supernova kinetic energy of 10$^{51}$ erg, we calculated an age assuming a Sedov model ($t_{\rm s}$) as

\begin{equation}
t_{\rm s} = 1.1~{\rm kyr} \,(n_{\rm e}/1~{\rm cm}^{-3})^{0.5} (D/ {\rm 5~pc})^{2.5}, 
\end{equation}
where $D$ is the SNR diameter \citep{sedov59}.

\subsection{Gamma-ray spectral analysis}\label{sec-gev}
In order to obtain gamma-ray properties with a systematic analysis, we extracted the gamma-ray spectra presented in the publications, and estimated the progress of the particle escape as follows.
We assume that hadronic emissions dominate the gamma-ray spectra.
The particle escape generally {develops} from higher particle energies, because of larger diffusion lengths (\cite{ptuskin03}).
Assuming that the maximum energies of accelerated protons are determined by particle escape, which should be natural for middle-aged to old SNRs, exponential cutoff energies ($E_{\rm cut}$) or break energies ($E_{\rm br}$) of the gamma-ray spectra represent the progress of the particle escape.
If we assume that escaping particles do not emit significant amount of gamma-rays, exponential-like cutoff features are expected. On the other hand, if the emission of escaping particles is seen as well, we expect the spectra which can be approximated with a broken power law \citep{ohira10}.

\begin{figure*}
\centering
\includegraphics[width=20cm, angle=90]{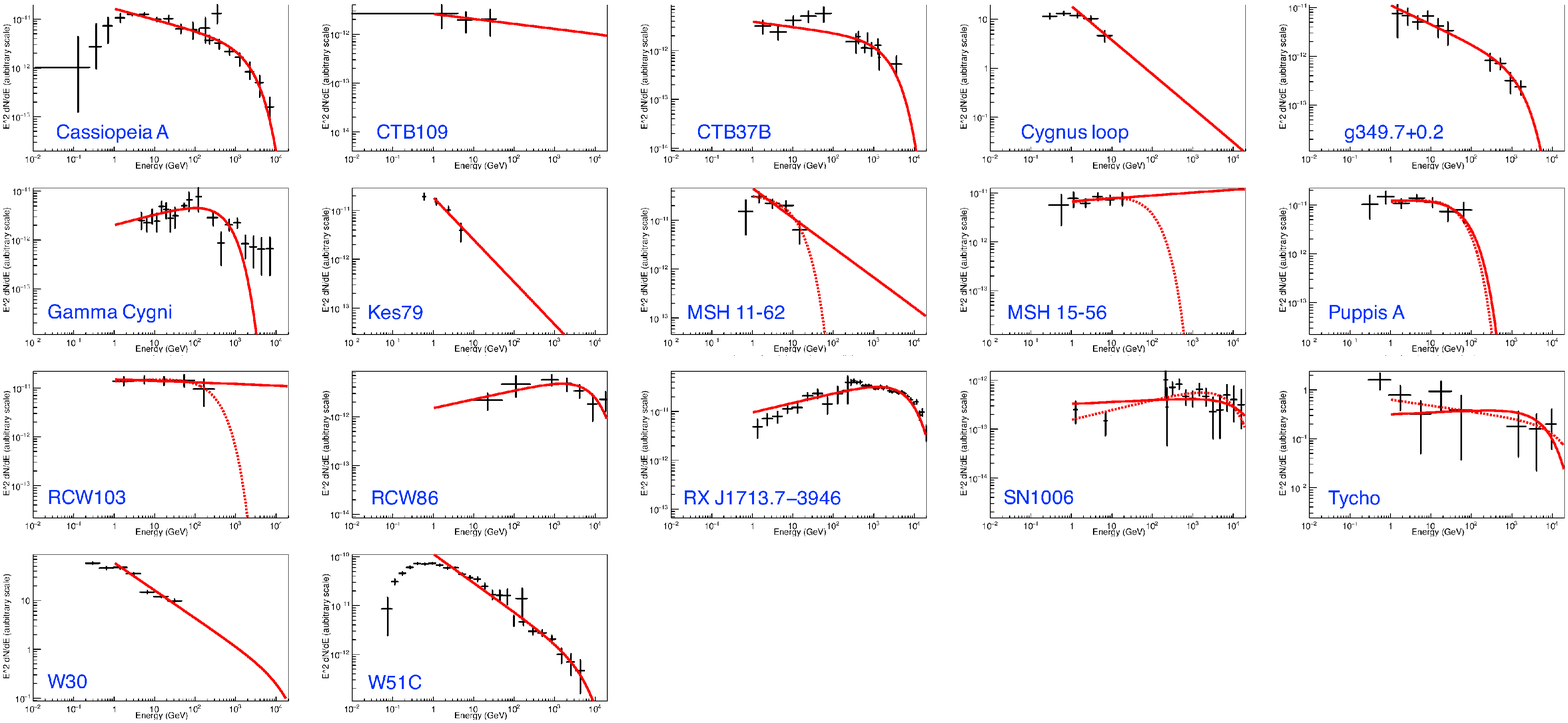}
\caption{Fitting results for the gamma-ray spectra with an exponential cutoff power law model. The dashed and solid lines represent the best-fitting models in cases (A) and (B), respectively (see text). The colors of the object names correspond to the plasma types (blue: IP; red: RP; black: CIE or no detection of thermal X-rays).}
\label{fig-gev-cut1}
\end{figure*}

\begin{figure*}
\centering
\includegraphics[width=20cm, angle=90]{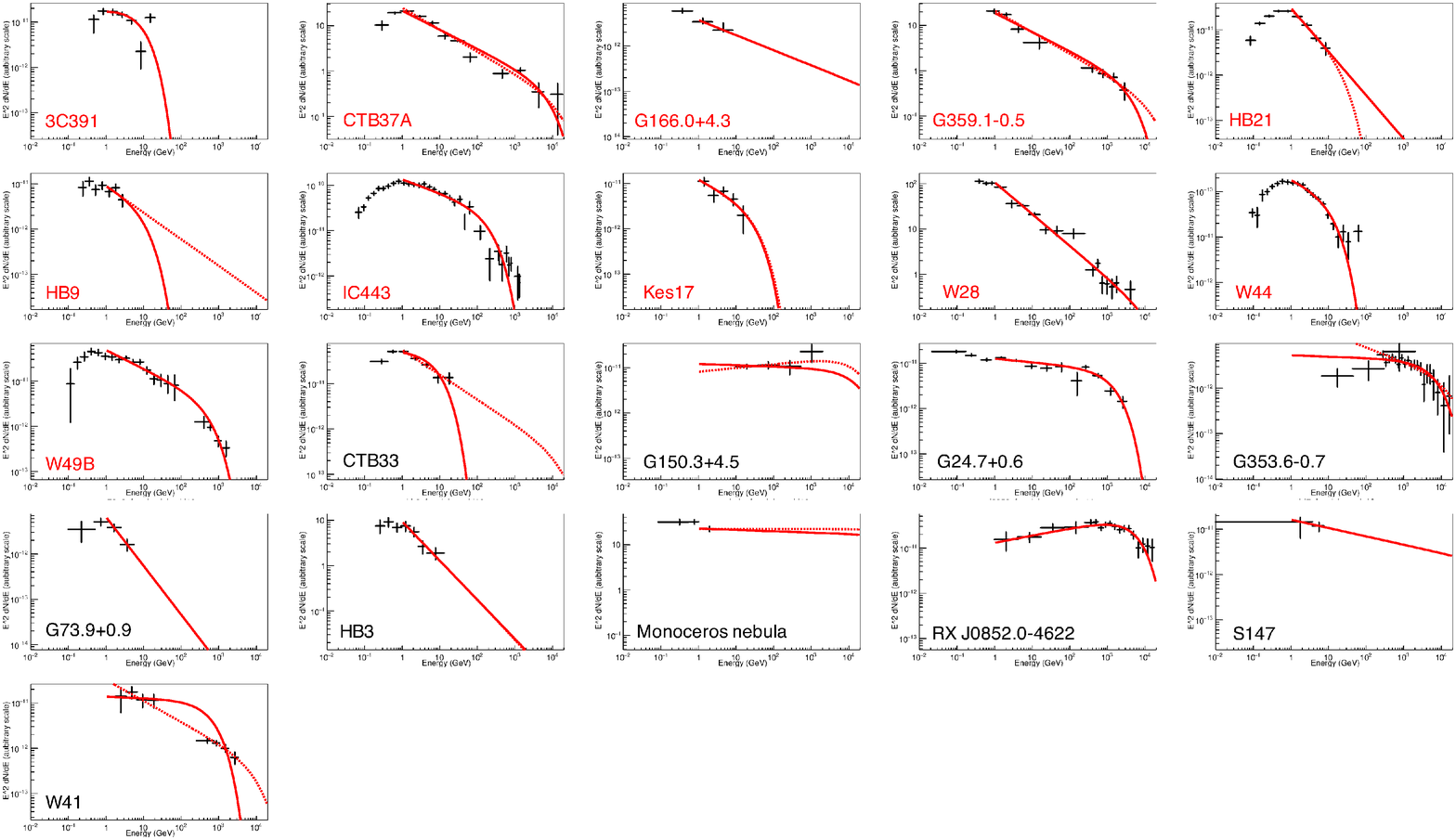}
\caption{Fitting results for the gamma-ray spectra with an exponential cutoff power law model. Same convention for lines is used as Figure \ref{fig-gev-cut1}. The colors of the object names correspond to the plasma types (blue: IP; red: RP; black: CIE or no detection of thermal X-rays).}
\label{fig-gev-cut2}
\end{figure*}

\begin{figure*}
\centering
\includegraphics[width=20cm, angle=90]{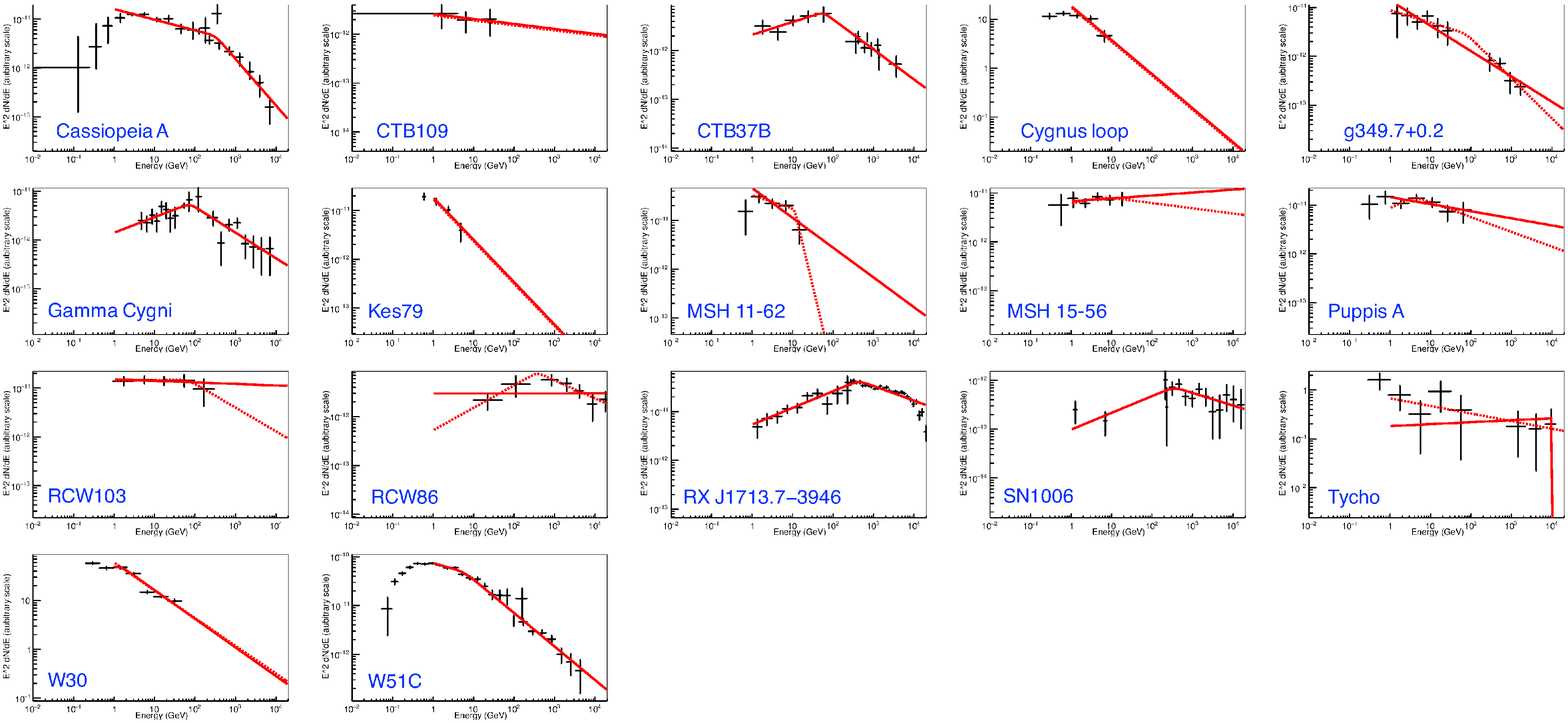}
\caption{Fitting results for the gamma-ray spectra with a broken power law model. Same conventions for lines and colors are used as Figure \ref{fig-gev-cut1}.}
\label{fig-gev-br1}
\end{figure*}

\begin{figure*}
\includegraphics[width=20cm, angle=90]{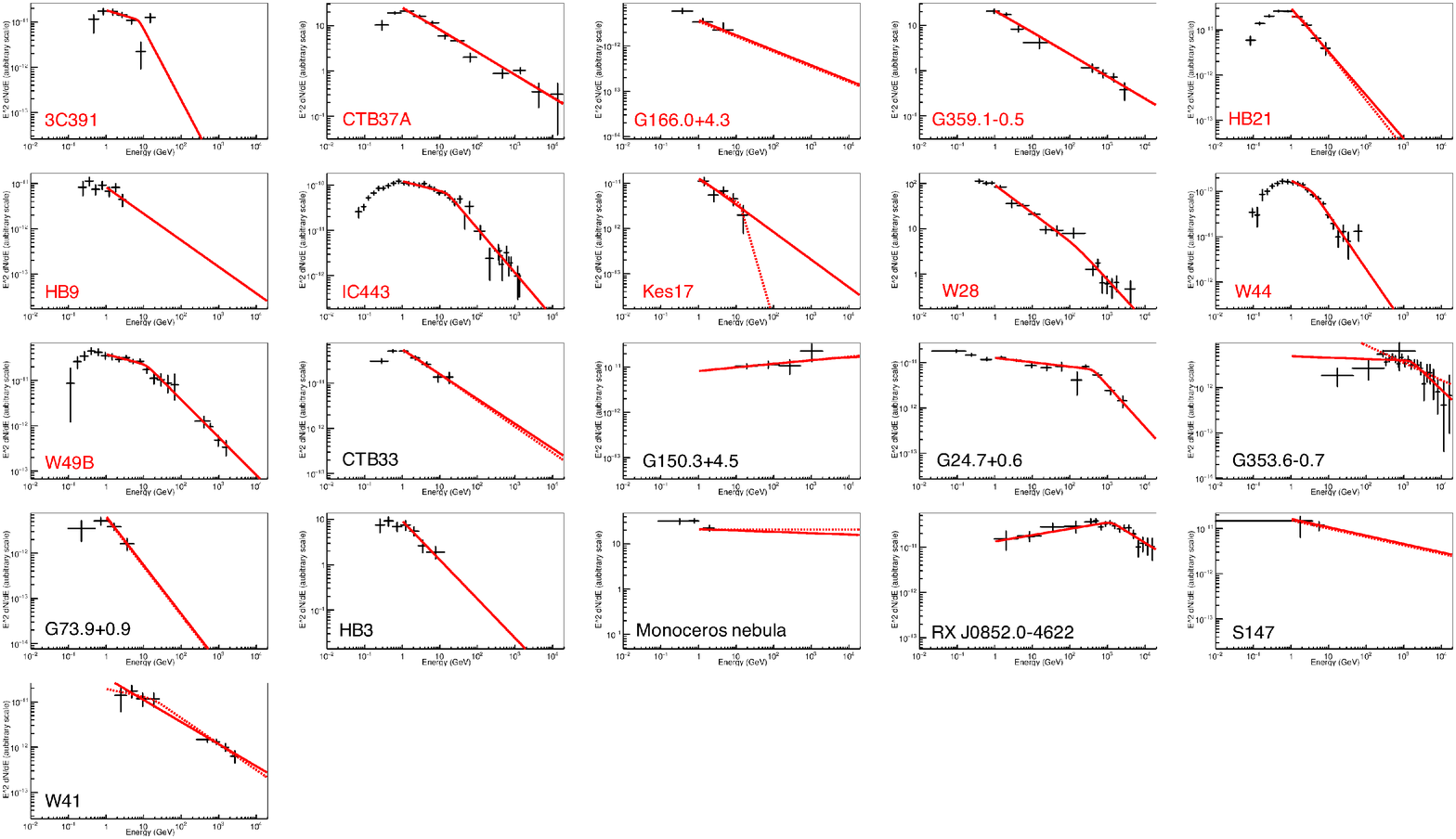}
\caption{Fitting results for the gamma-ray spectra with a broken power law model. Same conventions for lines and colors are used as Figure \ref{fig-gev-cut1}.}
\label{fig-gev-br2}
\end{figure*}

Thus, we fitted the gamma-ray spectra with both an exponential cutoff power law model,
\begin{equation}
E^{2}\,\frac{dN}{dE} = {\rm A}\times E^{-\Gamma_{\rm cut}} {\rm exp}(-E/E_{\rm cut}),
\end{equation}
and a (smoothed) broken power law model,
\begin{equation}
E^{2}\,\frac{dN}{dE} = {\rm A}\times E^{-\Gamma_{\rm br}} (1 + (E/E_{\rm br})^\frac{-\Gamma_{\rm br} - \Gamma_{\rm br2}}{0.1})^{0.1}.
\end{equation}
In ether case, first we fitted the spectra with a model which has free parameters of the normalization (A), photon indices ($\Gamma_{\rm cut}, \Gamma_{\rm br}$ and $\Gamma_{\rm br2}$) and the cutoff energy ($E_{\rm cut}$) or break energy ($E_{\rm br}$) (case (A)). The fitting results are shown in Figure \ref{fig-gev-cut1} to \ref{fig-gev-br2}.
The dashed red lines represent the best-fitting spectra in the case (A).
While the photon indices were well constrained, the cutoff energies were not determined for most of the sources (see Figure \ref{fig-di} (b) and \ref{fig-di-br} (b)).
The average values of the $\Gamma_{\rm cut}$of the IP SNRs, RP SNRs, and the others obtained with the exponential cutoff power law models are 1.96$\pm$0.07, 2.44$\pm$0.05, and 2.03$\pm$0.07, respectively.
Those obtained with the broken power law models ($\Gamma_{\rm br}$) are 1.93$\pm$0.07, 2.38$\pm$0.05, and 2.02$\pm$0.05, all of which are consistent with $\Gamma_{\rm cut}$.
Thus, we modified the fitting algorithm as follows (case (B)): if the upper limit of the $E_{\rm cut}$ or $E_{\rm br}$ in case (A) can not be determined, we then fix the photon index to the average value of its group (IP, RP or others) and accept the result if $\chi^2$/d.o.f $<$ 2.0.
{In order to avoid overfitting, however, we did not use $E_{\rm cut}$ or $E_{\rm br}$ obtained if the $\chi^2$/d.o.f. is too small ($< 0.3$), and a simple power law fitting result was used instead in either case of (A) or (B), and is presented in Figure \ref{fig-gev-cut1} to \ref{fig-gev-br2}.}
The best-fitting models in case (B) are shown in Figure \ref{fig-gev-cut1} to \ref{fig-gev-br2} with the solid red lines.
For the panels in which only the solid red lines are shown, they represent the best-fitting models in case (A).
The best-fitting parameters are presented in Table \ref{tab-results}.

\clearpage
\onecolumn
\fontsize{8}{9}\selectfont
\begin{landscape}
\centering
\begin{ThreePartTable}
\begin{TableNotes}
\item[*]\label{tn-a} The ambient gas density.
\item[\dag]\label{tn-b} {References for the physical parameters. The references for $D$; $n_{\rm e}$ and $t_{\rm p}$; gamma-ray spectrum and $n_{\rm gas}$; distance are presented:}
(1) \citet{green17};
(2) \citet{murray79}; (3) \citet{ahnen17};
(4) \citet{reed95};
(5) \citet{sasaki13}; (6) \citet{castro12};
(7) \citet{kothes02};
(8) \citet{aharonian08}; (9) \citet{xin16};
(10) \citet{caswell75};
(11) \citet{zhou10}; (12) \citet{katagiri11};
(13) \citet{blair05};
(14) \citet{slane02}; (15) \citet{ergin15}; (16) \citet{hess15b}
(17) \citet{tian14};
(18) \citet{hui15}; (19) \citet{fraija16}; (20) \citet{aliu13};
(21) \citet{higgs77}; (22) \citet{lozinskaya00};
(23) \citet{giacani09}; (24) \citet{sun04}; (25) \citet{auchettl14};
(26) \citet{case98};
(27) \citet{slane12};
(28) \citet{moffett01}; (29) \citet{moffett02};
(30) \citet{cesur19}; (31) \citet{temim13};
(32) \citet{rosado96};
(33) \citet{petre82}; (34) \citet{hess15c};
(35) \citet{reynoso03};
(36) \citet{braun19}; (37) \citet{xing14};
(38) \citet{carter97};
(39) \citet{lemoine12}; (40) \citet{yuan14};
(41) \citet{rosado96}; (42) \citet{sollerman03};
(43) \citet{katsuda15}; (44) \citet{hess18};
(45) \citet{fukui03};
(46) \citet{yamaguchi08}; (47) \citet{condon17};
(48) \citet{winkler03};
(49) \citet{hwang02}; (50) \citet{giordano12}; (51) \citet{acciari11};
(52) \citet{tian11}; (53) \citet{hayato10};
(54) \citet{finley94}; (55) \citet{ajello12};
(56) \citet{kassim90}; (57) \citet{brand93};
(58) \citet{sasaki14}; (59) \citet{jogler16};
(60) \citet{koo97a}; (61) \citet{koo97b}; (62) \citet{green97};
%
%
(63) \citet{ergin14}; (64) \citet{sato14};
(65) \citet{radhak72};
(66) \citet{yamauchi14}; (67) \citet{abdollahi17};
(68) \citet{tian12};
(69) \citet{matsumura17}; (70) \citet{araya13};
(71) \citet{landecker82};
(72) \citet{suzuki20}; (73) \citet{hui16};
(74) \citet{suzuki18}; (75) \citet{ambrogi19};
(76) \citet{byun06};
(77) \citet{sezer19};
(78) \citet{leahy07};
(79) \citet{matsumura18}; (80) \citet{ackermann13};
(81) \citet{welsh03};
(82) \citet{washino16}; (83) \citet{gelfand13};
(84) \citet{caswell75};
(85) \citet{okon18}; (86) \citet{cui18};
(87) \citet{velazquez02};
(88) \citet{uchida12};
(89) \citet{seta98};
(90) \citet{hess18};
(91) \citet{moffett94};
%
%
(92) \citet{castro13};
(93) \citet{sarma97};
(94) \citet{ackermann17};
(95) \citet{cohen16};
(96) \citet{magic19};
(97) \citet{reich84}; (98) \citet{leahy89};
(99) \citet{doroshenko17}; (100) \citet{condon17};
(101) \citet{tian08}; (102) \citet{hess11};
(103) \citet{zdziarski16};
(104) \citet{pavlovic13};
(105) \citet{lazendic06}; (106) \citet{katagiri16a};
(107) \citet{routledge91};
(108) \citet{leahy86}; (109) \citet{katagiri16b};
(110) \citet{turner76}; (111) \citet{odegard86};
(112) \citet{slane01}; (113) \citet{aharonian07}; (114) \citet{tanaka11};
(115) \citet{katsuda08};
(116) \citet{katsuta12};
(117) \citet{chatterjee09}; (118) \citet{cordes02};
(119) \citet{misanovic11}; (120) \citet{hess15a}; 
(121) \citet{leahy08}
\end{TableNotes}

\begin{longtable}{*{7}{l}}
\caption{Properties of the SNRs used in this study.} \label{tab-snrs}\\
\hline\hline
Name & $D$ (pc) & Distance (kpc) & $n_{\rm e}$ (cm$^{-3}$) & $n_{\rm gas}$ (cm$^{-3}$)\tnotex{tn-a} & $t_{\rm p}$ (kyr)  & {Refs.\tnotex{tn-b}}  \\
\hline
\endfirsthead
\hline\hline
Name & $D$ (pc) & Distance (kpc) & $n_{\rm e}$ (cm$^{-3}$) & $n_{\rm gas}$ (cm$^{-3}$)\tnotex{tn-a} & $t_{\rm p}$ (kyr)  & {Refs.\tnotex{tn-b}}  \\
\hline
\endhead
\endfoot
\hline
\insertTableNotes
\endlastfoot

IP SNRs & & & & & & \\
Cassiopeia A (G111.7$-$.1) & 4.94 & 3.4 & 6 & 1 & 1.1 & 1, 4; 2; 3; 4 \\
CTB 109 (G109.1$-$1.0) & 24.4  & 3.1 & 0.9 & 1.1 & 17.7 (3.5--35) & 1, 7; 5; 6; 7                \\
CTB 37 B (G348.7+0.3) & 40  & 13.2 & 2 & 10 & 1.1 (0.63--2.54) & 1, 10; 8; 9; 10               \\
Cygnus loop (G74.0$-$8.5) & 30.7 (25.2--36.2) & 0.54 & 2 & 5 & 12.7 & 1, 13; 11; 12; 13                  \\
G349.7+0.2 & 7.53 (6.7--8.36) & 11.5 & 4.2 & 35 & 1.29 (1.06--1.67) & 1, 16; 14; 15, 16; 17                 \\
Gamma Cygni (G78.2+2.1) & 26.2  & 2.0 & 0.24 & 2 & 3.18 & 1, 21, 22; 18; 19, 20; 21, 22       \\
Kes 79 (G033.6+0.1)	& 19.2  & 7.0 & 1 & 3 & 1.97 (1.91--2.04) & 1, 26; 23, 24; 25; 26      \\
MSH 11-62 (G291.0$-$0.1)	& 20.4 (18.9--21.8) & 5 & 0.16 & 6.8 & 2.99 (2.59--3.38) & 1, 28, 29; 27; 27; 28, 29                \\
MSH 15-56 (G326.3$-$01.8)	& 45.3  & 4.1 & 0.15 & 1 & 29.7 (25.5--34) & 1, 32; 30; 31; 32            \\
Puppis A (G260.4$-$3.4)	& 35.2 (32.0--38.4) & 2.2 & 1 & 4 & 7.42 (7.07--7.9) & 1, 35; 33; 34; 35                \\
RCW 103 (G332.4$-$0.4)	& 10  & 3.3 & 5.7 & 10 & 3.41 & 1, 38; 36; 37; 38         \\
RCW 86 (G315.4$-$2.3)	& 30  & 2.5 & 2 & 1 & 1.1 (0.32--1.86) & 1, 42; 39; 40; 41, 42       \\
RX J1713.7-3946 (G347.3$-$0.5)	& 17.5 (16.0--18.9) & 1 & 0.1 & 0.01 & 159 (127--191) & 1, 45; 43; 44; 45                 \\
SN 1006 (G327.6+14.6)	& 18  & 2.2 & 0.15 & 0.085 & 5.25 (4.5--5.99) & 1, 48; 46; 47; 48                 \\
Tycho (G120.1+1.4)	& 6.6  & 3 & 0.13 & 10 & $<$ 0.073 & 1, 51, 52; 49; 50, 51; 52, 53           \\
W 30 (G8.7$-$0.1)	& 52.4  & 4 & 0.15 & 100 & 276 & 1, 56, 57; 54; 55; 56, 57            \\
W 51 C (G49.2$-$0.7)	& 48  & 4.3 & 0.07 & 10 & 123 (61.4--187) & 1, 60--62; 58; 59; 60--62            \\

&&&&&&\\
RP SNRs & & & & & & \\
3C 391 (G31.9+0.0)	& 18.5 (16--21) & 7.2 & 0.9 & 300 & 45 (38--49) & 1, 65; 63, 64; 63; 65  \\
CTB 37 A (G348.5+0.1)	& 44  & 7.9 & 0.8 & 100 & 52 (48--64) & 1, 68; 66; 67; 68 \\
G166.0+4.3	& 65.5 (51--80) & 4.5 & 0.9 & 0.01 & 69 (65--75) & 1, 71; 69; 70; 71 \\
G359.1$-$0.5	& 28  & 4.6 & 0.7 & 100 & 19 (17--21) & 1, 72; 72; 73; 72  \\
HB 21 (G89.0+4.7)	& 52.5 (45--60) & 1.7 & 0.06 & 15 & 170 (110--250) & 1, 76; 74; 75; 76  \\
HB 9 (G160.9+2.6)	& 30  & 0.8 & 0.9 & 0.1 & 20.5 (19.1--21.6) & 1, 78; 77; 77; 78 \\
IC 443 (G189.1+3.0)	& 20  & 1.5 & 1.6 & 140 & 12 (11--13) & 1, 81; 79; 80; 81    \\
Kes 17 (G304.6+0.1)	& 35  & 10 & 0.9 & 10 & 57 (46--78) & 1, 84; 82; 83; 84   \\
W 28 (G6.4$-$0.1)	& 28  & 2 & 1 & 100 & 35 (32--41) & 1, 87; 85; 86; 87  \\
W 44 (G34.7$-$0.4)	& 27.5 (24--31) & 3 & 1 & 200 & 20 (18--23) & 1, 89; 88; 80; 89  \\
W 49 B (G43.3$-$0.2)	& 8 (7--9) & 10 & 2.7 & 700 & 5.2 (4.7--5.7) & 1, 91; 79; 90; 91  \\

&&&&&&\\
Others & & & & & & \\
CTB 33 (G337.0$-$0.1)	& 5.1  & 11 & --- & 60 & --- & 1, 93; ---; 92; 93  \\
G150.3+4.5	& 18.8  & 0.4 & --- & 1 & --- & 1, 95; ---; 94; 95   \\
G24.7+0.6	& 32.7 (21.8--43.6) & 5 & --- & --- & --- & 1, 98; ---; 96; 97, 98  \\
G353.6$-$0.7	& 28  & 3.2 & --- & 0.01 & --- & 1, 101, 102; 99; 100; 101, 102  \\
G73.9+0.9	& 32  & 4 & --- & 10 & --- & 1, 104; ---; 103; 104  \\
HB 3 (G132.7+1.3)	& 52.8  & 2.2 & 0.32 & 2 & --- & 1, 107; 105; 106; 107 \\
Monoceros nebula (G205.5+0.5)	& 102.5  & 2 & 0.003 & 3.6 & --- & 1, 110, 111; 108; 109; 110, 111  \\
Vela Jr. (RX J0852.0-4622; G266.2$-$1.2)		& 6.98  & 0.75 & 0.03 & 2 & --- & 1, 115; 112; 113, 114; 115  \\
S 147 (G180.0$-$1.7)	& 68.1  & 1.3 & --- & 250 & --- & 1, 117, 118; ---; 116; 117, 118  \\
W 41 (G23.3$-$0.3)	& 33  & 4.2 & --- & 4 & --- & 1, 121; 119; 120; 121   \\

\end{longtable}
\end{ThreePartTable}
\end{landscape}
\normalsize

\fontsize{8}{9}\selectfont
\begin{landscape}
\centering
\begin{ThreePartTable}
\begin{TableNotes}
\item[*]\label{tn-a2} Best-fitting $\Gamma_{\rm cut}$ in case (A).
\item[\dag]\label{tn-b2} Best-fitting $\Gamma_{\rm br}$ in case (A).
\item[\ddag]\label{tn-c2} Values are not presented if the cutoff/break energies are not obtained.
\item[\S]\label{tn-d2} Luminosity in 1--100 GeV energy range calculated from the best-fitting parameters of the exponential cutoff power law model in case (B).
\item[$\|$]\label{tn-e2} Luminosity in 1 GeV to 20 TeV energy range normalized at 1 GeV calculated from the best-fitting parameters of the exponential cutoff power law model in case (B).
\item[\#]\label{tn-f2} Ratio of the 10 GeV to 20 TeV and the 1--10 GeV luminosities, calculated using the best-fitting parameters of the exponential cutoff power law model in case (B).
\item[**]\label{tn-g2} Values are not constrained due to the large uncertainties of the spectral data.
\end{TableNotes}

\begin{longtable}{*{8}{l}}
\caption{GeV--TeV spectral fitting results.} \label{tab-results}\\
\hline\hline
Name & $\Gamma_{\rm cut}$ \tnotex{tn-a2} & $\Gamma_{\rm br}$ \tnotex{tn-b2}   & $E_{\rm cut}$ (GeV)\tnotex{tn-c2} & $E_{\rm br}$ (GeV)\tnotex{tn-c2} & L$_{\rm1\mathchar`-100 \,GeV}$\tnotex{tn-d2} & $\hat{L}$ \tnotex{tn-e2} & $R_{\rm GeV}$ \tnotex{tn-f2}\\ 
\hline
\endfirsthead
\hline\hline
Name & $\Gamma_{\rm cut}$ \tnotex{tn-a2} & $\Gamma_{\rm br}$ \tnotex{tn-b2}   & $E_{\rm cut}$ (GeV)\tnotex{tn-c2} & $E_{\rm br}$ (GeV)\tnotex{tn-c2} & L$_{\rm1\mathchar`-100 \,GeV}$\tnotex{tn-d2} & $\hat{L}$ \tnotex{tn-e2} & $R_{\rm GeV}$ \tnotex{tn-f2} \\ 
\hline
\endhead
\endfoot
\hline
\insertTableNotes
\endlastfoot

IP SNRs\\
Cassiopeia A (G111.7$-$2.1)	& 2.22 (2.2--2.25) & 2.22 (2.19--2.24) & 2.1 (1.6--2.8)$\times10^{3}$ & 3.2 (1.4--4.8)$\times10^{2}$ & 1.73 (1.71--1.75)$\times 10^{34}$ & 3.18 (2.85--3.5) & 0.94 (0.84--1.04) \\
CTB 109 (G109.1$-$1.0)	& 2.1 (1.8--2.4) & 2.1 (1.8--2.4) & --- & --- & 2.95 (1.71--4.18)$\times 10^{33}$ & 4.09 (2.06--6.12) & 1.36 (0.58--2.15) \\
CTB 37A (G348.5+0.1)	& 2.11 (2.08--2.15) & 1.74 (1.62--1.86) & 2.2 (1.1--5.5)$\times10^{3}$ & 58 (23--100) & 7.83 (7.07--8.59)$\times 10^{34}$ & 4.44 (3.36--5.53) & 1.34 (1.03--1.65) \\
Cygnus loop (G074.0$-$8.5)	& 2.70 (2.26--3.14) & 2.7 (2.26--3.14) & --- & --- & 3.88 (3.12--4.65)$\times 10^{32}$ & 1.24 (0.863--1.62) & 0.16 (0.04--0.281) \\
G349.7+0.2	& 2.39 (2.34--2.45) & 2.25 (2.09--2.42) & 1.5 (0.8--5.6)$\times10^{3}$ & --- & 10.0 (9.2--10.8)$\times 10^{34}$ & 2.10 (1.57--2.63) & 0.54 (0.38--0.71) \\
Gamma-cygni (G078.2+2.1)	& 1.78 (1.73--1.83) & 1.68 (1.59--1.78) & 4.7 (3.3--8.9)$\times10^{2}$ & 74 (46--120) & 3.04 (2.99--3.09)$\times 10^{33}$ & 10.8 (5.91--15.7) & 2.68 (1.72--3.65) \\
Kes 79 (G033.6+0.1)	& 2.87 (2.63--3.11) & 2.87 (2.63--3.11) & --- & --- & 3.32 (2.75--3.89)$\times 10^{34}$ & 0.86 (0.71--1.01) & 0.09 (0.006--0.18) \\
MSH 11-62 (G291.0$-$0.1)	& 2 (1.75--2.25) & 2.27 (2.03--2.52) & --- & --- & 1.28 (1.08--1.48)$\times 10^{35}$ & 1.35 (0.85--1.84) & 0.243 (0.127--0.359) \\
MSH 11-56 (G326.3$-$01.8)	& 1.87 (1.73--2) & 1.82 (1.55--2.09) & --- & --- & 2.22 (1.64--2.80)$\times 10^{34}$ & 8.86 (4.34--13.4) & 2.86 (1.07--4.64) \\
Puppis A (G260.4$-$3.4)	& 1.89 (1.78--2.01) & 1.62 (1.23--2.00) & 63 (36--140) & --- & 7.28 (6.06--8.50)$\times 10^{33}$ & 3.98 (3.24--4.72) & 0.81 (0.53--1.09) \\
RCW 103 (G332.4$-$0.4)	& 1.95 (1.87--2.03) & 1.99 (1.9--2.08) & --- & --- & 2.46 (2.26--2.66)$\times 10^{34}$ & 6.42 (4.47--8.37) & 2.22 (1.49--2.96) \\
RCW 86 (G315.4$-$2.3)	& 1.82 (1.75--1.89) & 1.54 (1.3--1.78) & 8.4 (4.8--17)$\times10^{3}$ & --- & 2.15 (1.89--2.41)$\times 10^{33}$ & 18.2 (9.14--27.2) & 5.50 (3.12--7.88) \\
RX J1713-3946 (G347.3$-$0.5)	& 1.8 (1.79--1.81) & 1.66 (1.63--1.68) & 6.4 (6.0--6.8)$\times10^{3}$ & 4.3 (3.6--6.5)$\times10^{2}$ & 2.28 (2.09--2.48)$\times 10^{33}$ & 18.6 (15.6--21.7) & 5.75 (4.94--6.57) \\
SN 1006 (G327.6+14.6)	& 1.8 (1.75--1.85) & 1.66 (1.55--1.76) & $>$2.0 & 3.4 (1.0--5.8)$\times10^{2}$ & 2.57 (2.34--2.81)$\times 10^{32}$ & 11.3 (10.9--11.8) & 3.90 (3.71--4.09) \\
Tycho (G120.1+1.4)	& 2.12 (2.02--2.23) & 2.16 (1.99--2.32) & $>$1.3 & $>$9.5 & 7.35 (4.01--10.7)$\times 10^{32}$ & 9.16 (7.50--10.8) & 2.97 (2.25--3.68) \\
W 30 (G008.7$-$0.1)	& 2.57 (2.48--2.66) & 2.57 (2.41--2.73) & $>$51 & $>$3.3 & 7.99 (7.04--8.94)$\times 10^{34}$ & 1.39 (1.15--1.62) & 0.29 (0.18--0.40) \\
W 51 C (G049.2$-$0.7)	& 2.6 (2.57--2.62) & 2.23 (2.17--2.29) & $>$2.7 & 5.9 (3.5--14) & 1.06 (1.02--1.11)$\times 10^{35}$ & 1.47 (1.36--1.58) & 0.339 (0.304--0.374) \\

\\
RP SNRs\\
3C 391 (G031.9+0.0)	& 1.9 (1.75--2.06) & 2.25 (2.07--2.43) & 7.4 (3.8--18) & 7.6 (3.3--9.4) & 5.64 (5.32--5.96)$\times 10^{34}$ & 1.71 (1.13--2.29) & 0.105 (0.052--0.159) \\
CTB 37 A (G348.5+0.1)	& 2.49 (2.45--2.52) & 2.49 (2.44--2.55) & $>$3.3 & $>$16 & 1.4 (1.3--1.5)$\times 10^{35}$ & 2.16 (2.13--2.19) & 0.602 (0.581--0.624) \\
G166.0+4.3	& 2.33 (1.92--2.75) & 2.33 (1.92--2.75) & --- & --- & 5.72 (3.87--7.58)$\times 10^{33}$ & 2.05 (1.33--2.77) & 0.577 (0.293--0.862) \\
G359.1$-$0.5	& 2.48 (2.44--2.51) & 2.49 (2.43--2.54) & 4.6 (2.6--11)$\times10^{3}$ & $>$53 & 2.59 (2.27--2.91)$\times 10^{34}$ & 2.15 (2.12--2.18) & 0.594 (0.573--0.615) \\
HB 21 (G089.0+4.7)	& 2.8 (2.54--3.05) & 1.96 (0.169--3.76) & --- & --- & 3.35 (2.58--4.12)$\times 10^{33}$ & 0.68 (0.44--0.91) & 0.084 (0.026--0.143) \\
HB 9 (G160.9+2.6)	& 2.59 (1.98--3.19) & 2.59 (1.98--3.19) & $>$1.2 & $>$3.5 & 2.32 (1.16--3.48)$\times 10^{32}$ & 1.36 (0.53--2.19) & 0.312 (0.0006--0.624) \\
IC 443 (G189.1+3.0)	& 2.3 (2.27--2.32) & 2.21 (2.18--2.25) & 2.1 (1.8--2.5)$\times10^{2}$ & 16 (11--22) & 2.30 (2.27--2.32)$\times 10^{34}$ & 2.24 (2.05--2.43) & 0.514 (0.470--0.557) \\
Kes 17 (G304.6+0.1)	& 2.44 (2.23--2.65) & 2.48 (2.25--2.72) & 30 (11--2400) & $>$1.0 & 6.24 (4.91--7.57)$\times 10^{34}$ & 1.66 (1.20--2.11) & 0.327 (0.086--0.568) \\
W 28 (G6.4$-$0.1)	& 2.7 (2.67--2.74) & 2.62 (2.56--2.69) & $>$2.0 & 1.5 (0.1--6.6)$\times10^{2}$ & 1.89 (1.59--2.19)$\times 10^{34}$ & 1.26 (1.11--1.42) & 0.263 (0.213--0.314) \\
W 44 (G034.7$-$0.4)	& 2.35 (2.28--2.42) & 2.42 (2.3--2.53) & 11 (6.5--20) & 3.8 (1.4--7.4) & 7.28 (7.19--7.38)$\times 10^{34}$ & 1.19 (1.03--1.36) & 0.083 (0.057--0.109) \\
W 49 B (G043.3$-$0.2)	& 2.41 (2.37--2.45) & 2.21 (2.13--2.3) & 5.7 (4.3--8.4)$\times10^{2}$ & 12 (3.5--16) & 3.14 (3.06--3.22)$\times 10^{35}$ & 1.89 (1.53--2.25) & 0.45 (0.34--0.56) \\

\\
Others\\
CTB 33 (G337.0$-$0.1)	& 2.55 (2.44--2.65) & 2.24 (1.12--3.37) & 8.0 (5.6--11) & --- & 3.37 (3.02--3.73)$\times 10^{35}$ & 1.66 (1.47--1.85) & 0.106 (0.066--0.146) \\
G150.3+4.5	& 1.92 (1.8--2.03) & 1.92 (1.76--2.08) & $>$2.7 & $>$1.0 & 2.67 (2.45--2.89)$\times 10^{32}$ & 7.39 (6.46--8.31) & 2.48 (2.05--2.92) \\
G24.7+0.6	& 2.08 (2.06--2.1) & 2.10 (2.08--2.12) & 1.5 (1.2--2.0)$\times10^{3}$ & 4.1 (2.9--5.7)$\times10^{2}$ & 3.84 (3.77--3.92)$\times 10^{34}$ & 4.73 (4.28--5.18) & 1.44 (1.31--1.56) \\
G353.6$-$0.7	& 2.32 (2.26--2.37) & 2.38 (2.31--2.45) & 6.8 (5.4--8.8)$\times10^{3}$ & 1.6 (1.2--2.1)$\times10^{3}$ & 7.50 (6.99--8.01)$\times 10^{33}$ & 7.17 (6.99--7.35) & 2.38 (2.30--2.47) \\
G73.9+0.9	& 3.08 (2.17--3.99) & 3.08 (2.17--3.99) & --- & --- & 3.10 (2.16--4.04)$\times 10^{33}$ & 0.605 (0.422--0.788) & 0.065 (0.022--0.108) \\
HB 3 (G132.7+1.3)	& 2.86 (2.54--3.19) & 2.86 (2.20--3.53) & --- & --- & 2.65 (2.47--2.83)$\times 10^{33}$ & 0.75 (0.48--1.02) & 0.107 (0.032--0.182) \\
Monoceros nebula (G205.5+0.5)\tnotex{tn-g2}	& --- & --- & --- & --- & --- & --- & --- \\
Vela Jr. (RX J0852.0-4622; G266.2$-$1.2)	& 1.84 (1.82--1.86) & 1.86 (1.84--1.88) & 5.4 (4.5--6.5)$\times10^{3}$ & 1.3 (0.9--1.6)$\times10^{3}$ & 4.07 (3.85--4.29)$\times 10^{34}$ & 14.8 (12.3--17.4) & 4.64 (3.97--5.31) \\
S 147 (G180.0$-$1.7)	& 2.18 (1.65--2.72) & 2.18 (1.65--2.72) & --- & --- & 2.74 (1.80--3.67)$\times 10^{33}$ & 3.17 (1.92--4.43) & 1.02 (0.52--1.51) \\
W 41 (G23.3$-$0.3)	& 2.46 (2.42--2.51) & 2.17 (1.91--2.44) & 6.3 (5.4--7.2)$\times10^{2}$ & --- & 3.23 (2.68--3.78)$\times 10^{34}$ & 5.25 (5.13--5.38) & 1.49 (1.43--1.55) \\


\end{longtable}
\end{ThreePartTable}
\end{landscape}
\normalsize
\twocolumn

\section{Systematic analysis on the time evolution of particle escape}\label{sec-sysanalysis}

\begin{figure*}
\includegraphics[width=16cm]{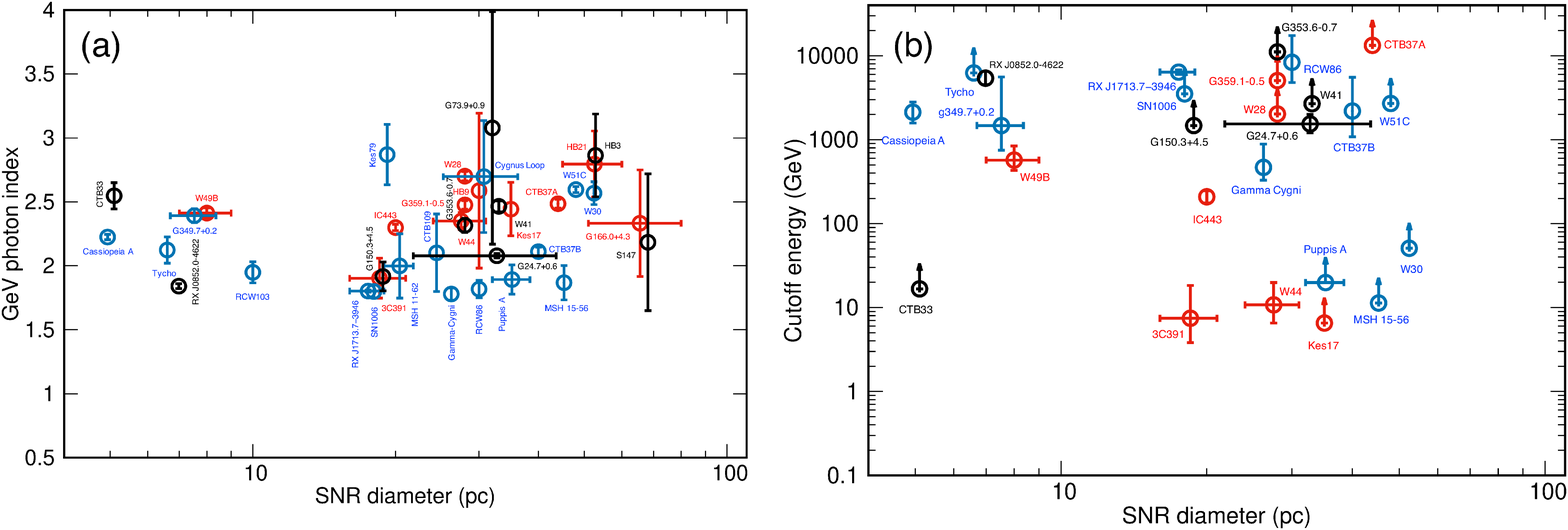}
\caption{Results obtained in the gamma-ray spectral fittings with an exponential cutoff power law model. (a) The relation between $D$ and $\Gamma_{\rm cut}$. (b) The relation between $D$ and $E_{\rm cut}$ obtained in the fittings with free photon indices. The colors of the data points and the names correspond to the plasma types (blue: IP; red: RP; black: CIE or no detection of thermal X-rays).
}
\label{fig-di}
\end{figure*}

\begin{figure*}
\includegraphics[width=16cm]{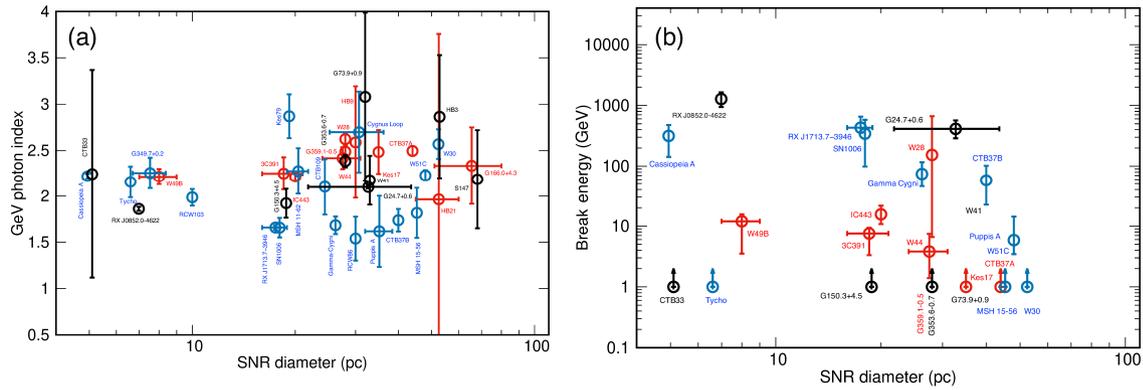}
\caption{Results obtained in the gamma-ray spectral fittings with a broken power law model. (a) The relation between $D$ and $\Gamma_{\rm br}$. (b) The relation between $D$ and $E_{\rm br}$ obtained in the fittings with free photon indices. Same conventions for colors of the data points and names are used as Figure \ref{fig-di}.
}
\label{fig-di-br}
\end{figure*}

Here, we make comparisons of the gamma-ray properties obtained in Section \ref{sec-data} and the thermal X-ray properties shown in Table \ref{tab-snrs}.
The distribution of the photon indices obtained in case (A) are shown in Figure \ref{fig-di} (a) and \ref{fig-di-br} (a). The objects with the RPs exhibit rather softer indices than the others (see Section \ref{sec-data}).

\begin{figure*}
\includegraphics[width=16cm]{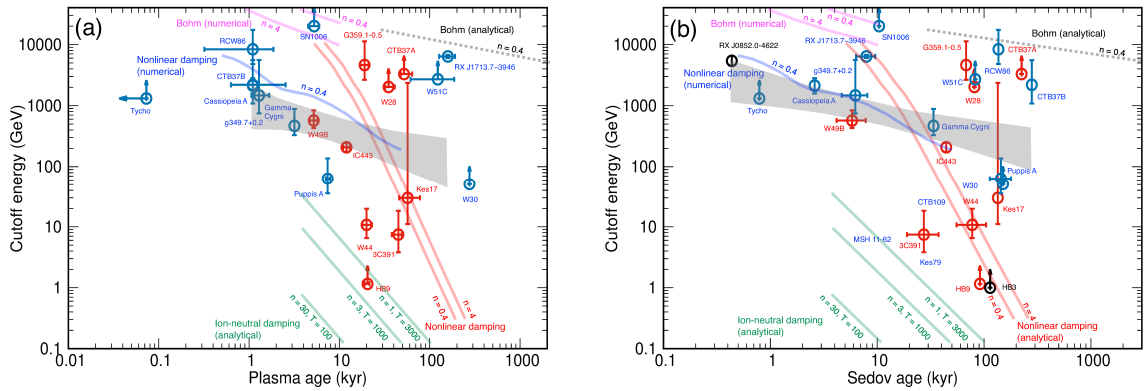}
\caption{(a) The relation between $t_{\rm p}$ and $E_{\rm  cut}$ obtained in the fittings with the model (B). (b) Same as (a) but for $t_{\rm s}$ instead of $t_{\rm p}$.
The gray regions represent 1 $\sigma$ confidence regions for the best-fitting power law function.
The dotted black, solid red and green lines represent the analytical models corresponding to the Bohm diffusion case, the case with nonlinear wave damping, and the case with ion-neutral wave damping, respectively, which are presented in \citet{ptuskin03}. The blue lines represent the numerical calculations in the Alfv\'enic diffusion (nonlinear wave damping) case presented in \citet{brose20}. The magenta lines represent the numerical calculations in the uniform ISM case by \citet{yasuda19}.
The $n$ and $T$ described along the model lines stand for the ambient density (cm$^{-3}$) and temperature (K), respectively. The ambient density refers to the cloud and the ISM density for ``ion-neutral damping'' case and for the other cases, respectively. In the case of ion-neutral damping, these values are set to keep a condition $nT = 3000$ K cm$^{-3}$ \citep{wolfire95}.
Same conventions as Figure \ref{fig-di} are used for the colors of the data points and names.
}
\label{fig-pc}
\end{figure*}

\begin{figure*}
\includegraphics[width=16cm]{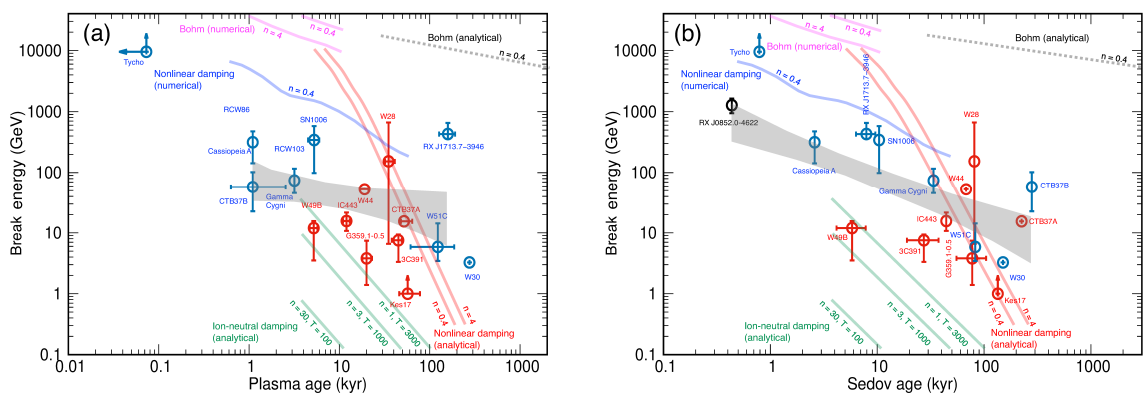}
\caption{(a) The relation between $t_{\rm p}$ and $E_{\rm br}$ obtained in the fittings with the model (B). (b) Same as (a) but for $t_{\rm s}$ instead of $t_{\rm p}$.
The gray regions represent 1 $\sigma$ confidence regions for the best-fitting power law function.
The theoretical model lines are the same as those in Figure \ref{fig-pc}.
Same conventions as Figure \ref{fig-di} are used for the colors of the data points and names.
}
\label{fig-pb}
\end{figure*}

Figure \ref{fig-pc} and \ref{fig-pb} show the {plots of the $E_{\rm cut}$ and $E_{\rm br}$ over} $t_{\rm p}$ and $t_{\rm s}$.
These are quite consistent with the result in \citet{zeng19}, which shows the plots of $E_{\rm br}$ over SNR ages.
{The gray regions are calculated by fitting the data with a power law model with free normalization and index.
They represent 1 $\sigma$ confidence regions for the best-fitting power law models, which show the trends of the data.}
{The correlation coefficient ($r$) between two parameters $x$ and $y$ is calculated as
\begin{equation} 
r = \frac{\frac{1}{n-1}\sum_{i=1}^n(x_i-\overline{x})(y_i-\overline{y})}{\sqrt{\frac{1}{n-1}\sum_{i=1}^n(x_i-\overline{x})^2}\sqrt{\frac{1}{n-1}\sum_{i=1}^n(y_i-\overline{y})^2}},
\end{equation}
where ${\overline{x}}$ and $\overline{y}$ stand for the average values, and $n$ and $i$ are the sample number and the sample index, respectively.
Here we note that, we exclude the data points with only given upper or lower limits in calculations of the correlation coefficients and the best-fitting functions.
We calculate a standard error of the regression as 
\begin{equation}
\sqrt{\frac{1-r^2}{n-2}},
\end{equation}
which is obtained based on Student's $t$ distribution with $(n-2)$ degrees of freedom.
Statistical errors are estimated as follows: we randomize the data points considering their uncertainties and calculate corresponding correlation coefficients for individual data sets, to obtain the distribution of the coefficients. Then we get the standard deviation of the distribution as the statistical error.}

The coefficients calculated in this work are summarized in Table \ref{tab-coef}.
{We are not able to find significant trends that $E_{\rm cut}$ falls with increasing $t_{\rm p}$ or $t_{\rm s}$. Also for $E_{\rm br}$, no clear trends are found (Table \ref{tab-coef}).}
If we exclude an outlier RX J1713.7-3946 (see Section \ref{subsec-1713}), however, the correlations become significant for most cases with $E_{\rm cut}$ and $E_{\rm br}$ (Table \ref{tab-coef}).
The dotted black, solid red and green lines are analytical models of the maximum energies of accelerated particles derived in \citet{ptuskin03}, which show the values in a Bohm limit, acceleration with wave damping by nonlinear wave-wave interactions caused by shock-ISM (interstellar medium) collisions, and acceleration with wave damping by shock-cloud collisions, respectively. The latter two cases include wave amplification by accelerated particles. In all the cases, an upstream magnetic field strength ($B_0$) of 5 $\mu$G is assumed.
This value has been found to be reasonable at least for SNRs which are older than 2 kyr (e.g., \cite{bamba05}).
In the case of ion-neutral damping, we consider slowing down of shocks in clouds as
\begin{equation}
v_{\rm c} \sim \frac{v_{0}}{1 + (n_{\rm c}/n_{\rm 0})^{0.5}}, 
\end{equation}
where $v_0$, $v_{\rm c}$, $n_{\rm 0}$ and $n_{\rm c}$ are shock velocities before and after the collision to clouds, and post-shock densities of the intercloud region (assumed to be 1 cm$^{-3}$) and the cloud, respectively \citep{chevalier99}.
The solid blue lines are numerical calculation results of the maximum energies in acceleration with Alfv\'enic diffusion (including nonlinear wave damping, thus the diffusion coefficient is time dependent), obtained by \citet{brose20}.
In this case, the maximum energies are calculated by fitting time-integrated particle spectra in the downstream of the shock with exponential cutoff power law models, as observational studies done in Section \ref{sec-gev}.
This simulation does only consider the resonant amplification of Alfv\'enic turbulence by the accelerated particles. Thus, the contribution of the turbulence to the background magnetic field of $B_0 = 5~\mu$G is neglected and the magnitude of the maximum energies may not be correct.
The solid magenta lines represent numerical calculations of the maximum energies in an uniform ISM case by \citet{yasuda19}. This simulation assumes a Bohm-like diffusion coefficient with $B_0 = 4~\mu$G but includes wave amplification by accelerated particles. In this case, maximum energies represent exact highest energies which freshly accelerated particles at each time can reach, and thus the magnitude of the maximum energies may be higher than the observational values.

\begin{figure*}
\includegraphics[width=16cm]{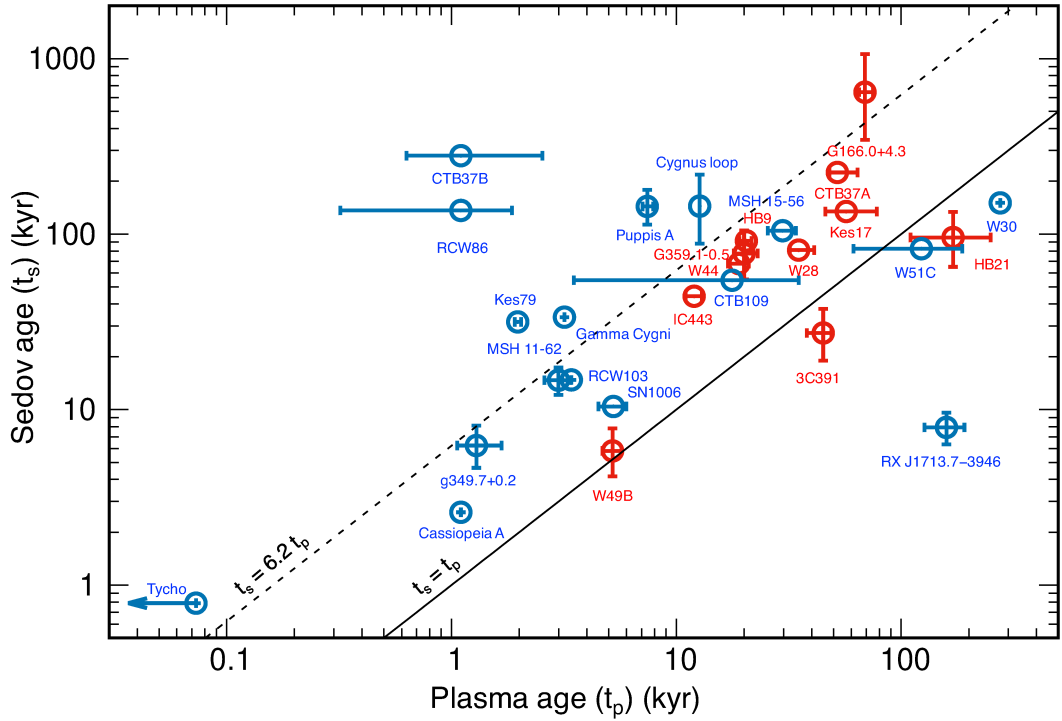}
\caption{The relation between $t_{\rm p}$ and $t_{\rm s}$. The solid and dashed black lines represent  linear functions $t_{\rm s} = t_{\rm p}$ and $t_{\rm s} = 6.2\, t_{\rm p}$, respectively.
Same conventions as Figure \ref{fig-di} are used for the colors of the data points and names.
}
\label{fig-ps}
\end{figure*}

Figure \ref{fig-ps} represents the relation between $t_{\rm p}$ and $t_{\rm s}$. The data show a clear correlation (Table \ref{tab-coef}).
The solid black line represents a linear function $t_{\rm s} = t_{\rm p}$. If we fit this trend with a linear function $t_{\rm s}/t_{\rm p} = f$, we get the best-fitting value of $f = 6.2 \pm 0.9$. This function is also presented in Figure \ref{fig-ps} with the dashed black line.
If we fit the IP SNRs and RP SNRs independently, we obtain the factors $f = 6.5 \pm 1.3$ and $3.6 \pm 0.4$, respectively.

\begin{figure*}
\includegraphics[width=16cm]{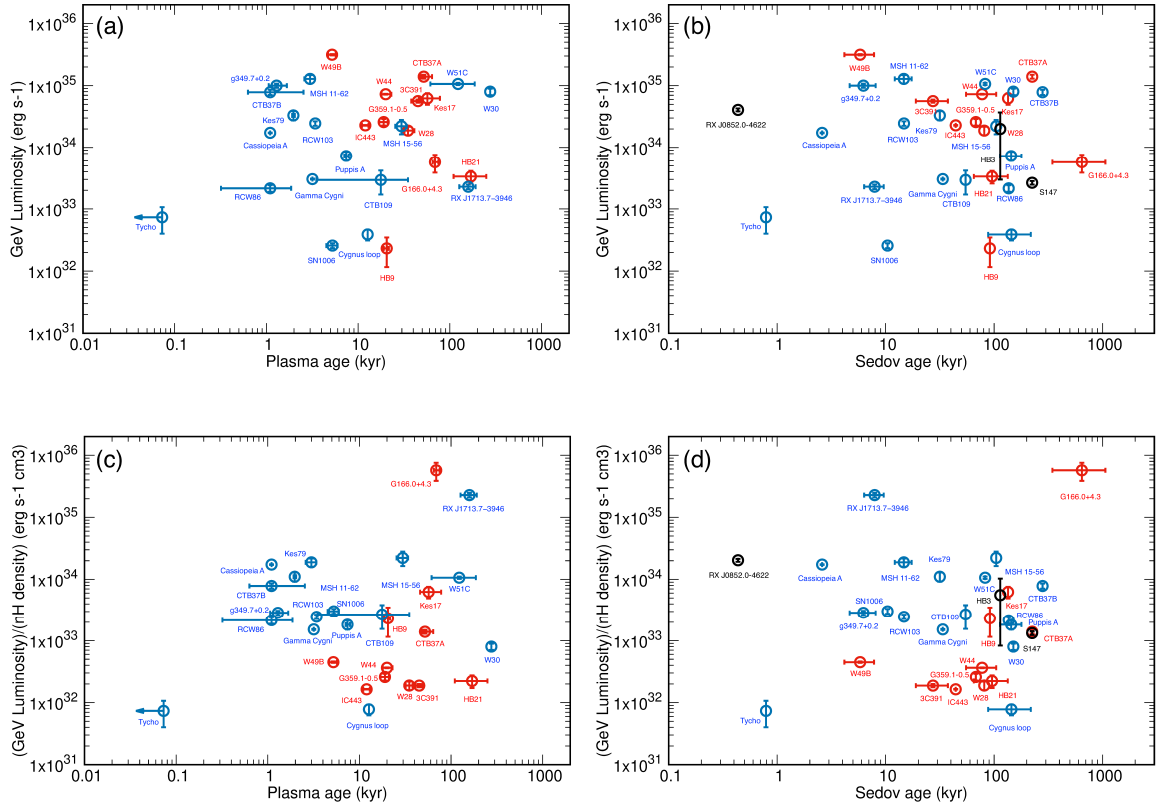}
\caption{(a) The {plots of the 1--100 GeV luminosity over} $t_{\rm p}$.
(b) Same as (a) but on $t_{\rm s}$.
(c) Dependency of the 1--100 GeV luminosity divided by the ambient gas density ($n_{\rm gas}$) on $t_{\rm p}$.
(d) Same as (c) but on $t_{\rm s}$.
Same conventions as Figure \ref{fig-di} are used for the colors of the data points and names.
}
\label{fig-pl}
\end{figure*}

In Figure \ref{fig-pl}, we show the {plots of the 1--100 GeV luminosity and that divided by the ambient proton density over} $t_{\rm p}$ and $t_{\rm s}$.
The luminosity is calculated from the best-fitting exponential cutoff power law model in case (B).
Although all the figures show no clear evolutions with $t_{\rm p}$ or $t_{\rm s}$, we can see that the objects with RPs have rather high luminosities than the others. This trend is consistent with the results presented in \citet{acero16}, which show that the 1--100 GeV luminosities of the SNRs of the ages above $\sim$ 10 kyr are $\sim$10$^{35}$ erg s$^{-1}$ and those of the younger ones are less than $\sim$10$^{35}$ erg s$^{-1}$.
We can also see that, when divided by the ambient density, the dispersion of the luminosity (logarithmic range from the minimum to the maximum value) becomes significantly narrow with a factor of $\sim$5, if RX J1713.7-3946 and G166.0+4.3 are excluded.

\begin{figure*}
\includegraphics[width=16cm]{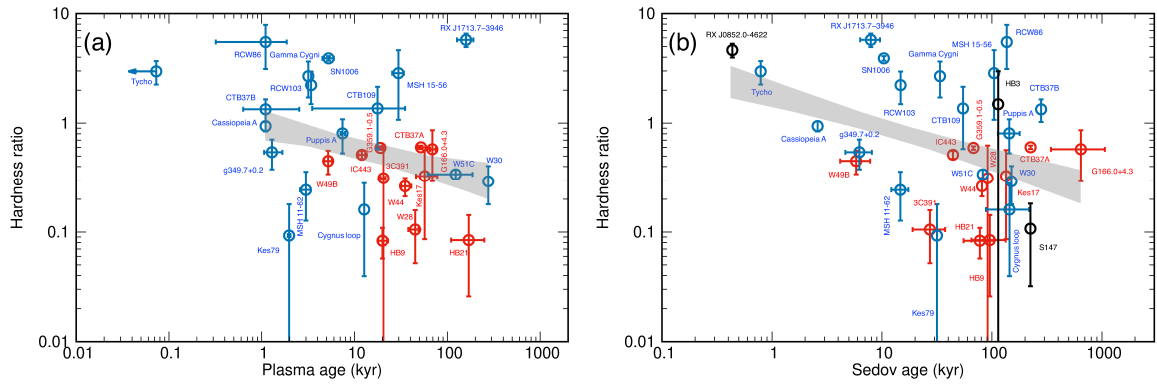}
\caption{(a) The {plots of the hardness ratio ($R_{\rm GeV}$) over} $t_{\rm p}$.
(b) Same as (a) but on $t_{\rm s}$.
Same conventions as Figure \ref{fig-di} are used for the colors of the data points and names.
}
\label{fig-ph}
\end{figure*}

\begin{figure*}
\includegraphics[width=16cm]{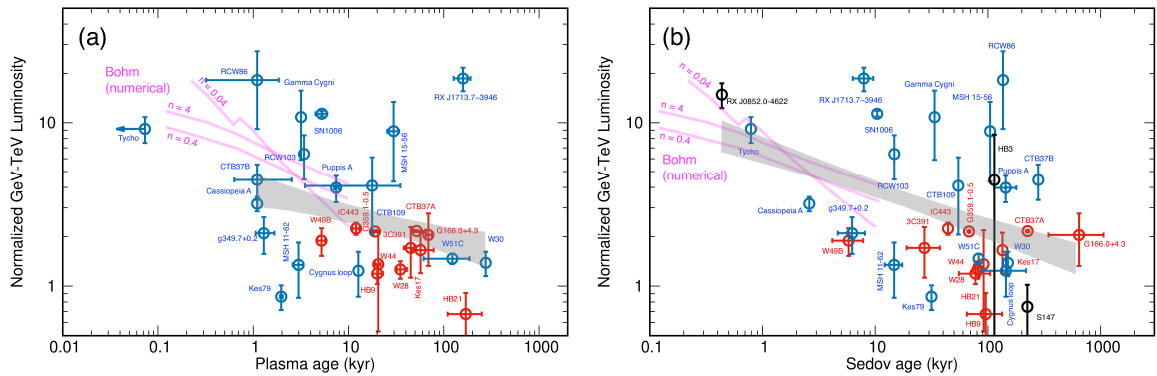}
\caption{(a) The {plots of the normalized luminosity ($\hat{L}$) over} $t_{\rm p}$.
(b) Same as (a) but on $t_{\rm s}$.
The gray regions represent 1 $\sigma$ confidence regions for the best-fitting power law model.
The magenta lines represent the numerical calculations in the uniform ISM case by \citet{yasuda19}.
Same conventions as Figure \ref{fig-di} are used for the colors of the data points and names.
}
\label{fig-psl}
\end{figure*}

In addition to $E_{\rm cut}$ and $E_{\rm br}$, we extract two more potential indicators of particle escape, namely a hardness ratio (ratio of the 10 GeV to 20 TeV and the 1--10 GeV luminosities; hereafter $R_{\rm GeV}$) and a normalized gamma-ray luminosity ($\hat{L}$). The $\hat{L}$ is the 1 GeV to 20 TeV luminosity normalized at 1 GeV, i.e.,
\begin{equation}
\hat{L} = \left(\int^{20~{\rm TeV}}_{1~{\rm GeV}} \frac{E^2\, dN\, (E)}{dE}\, dE\right)\, / \frac{E^2\, dN\, (1~ {\rm GeV})}{dE}.
\end{equation}
These are less biased values since they are independent of spectral modelings.
Figure \ref{fig-ph} shows the results for the ${R_{\rm GeV}}$. Comparison to either $t_{\rm s}$ or $t_{\rm p}$ shows a good trend (Table \ref{tab-coef}).
Figure \ref{fig-psl} shows the results for $\hat{L}$. Either compared to $t_{\rm p}$ or $t_{\rm s}$, $\hat{L}$ shows good trends (Table \ref{tab-coef}).
The gray regions represent 1 $\sigma$ confidence regions of the best-fitting power law models as in Figure \ref{fig-pc} and \ref{fig-pb}.
The solid magenta lines represent the numerical calculations in the uniform ISM case by \citet{yasuda19}. The $\hat{L}$ is calculated for the freshly accelerated particles at each time.

The objects with hard $\Gamma_{\rm cut}$ or $\Gamma_{\rm br}$ (significantly less than 2.0, i.e., CTB 37 A; Gamma Cygni; RCW 86; RX J1713.7-3946; SN 1006; Vela Jr.) are possibly emitting gamma-rays via inverse-Compton (IC) scatterings \citep{ohira12}. Thus they might be inapplicable in this study, since we assume that dominant component of the gamma-ray spectra is hadronic emission.
In particular, the emission mechanism of RX J1713.7-3946 has long been discussed and yet to be confirmed \citep{abdo10, ellison10, inoue12, gabici14, ohira17}.
Therefore, we calculate the correlation coefficients without these objects as well. The results are shown in Table \ref{tab-coef}.
Although the differences between the coefficients calculated with and without these objects are insignificant, all of the latter cases show significant correlations for $t_{\rm p}$.
{Oppositely, no significant correlations are found for $t_{\rm s}$ in this sample selection.}

{Table \ref{tab-coef2} shows the correlation coefficients for each of the IP and RP SNRs, to search for differences between their behavior.
Most of the coefficient values are not significant, and are consistent with each other for individual cases.
Thus, almost no significant differences are found. In the case of $t_{\rm p}$ vs. $t_{\rm s}$, we confirm that each of the IP and RP SNRs has a significant correlation if RX J1713.7-3946 is excluded.}

{We note that the correlation coefficients obtained above are subject to potential uncertainties of distances and $n_{\rm e}$.
The uncertainties of the distances have not been estimated quantitatively for most of the sample. These potential uncertainties might decrease the coefficients obtained above.
The uncertainties of $n_{\rm e}$ are also hard to estimate due to unknown values of the filling factors of the plasmas.
Here, in order to check this effect, we assign an error of 50\% to individual $n_{\rm e}$ values as an example and calculate all the correlation coefficients as above.
The resultant coefficients do not change much, with only the statistical errors increased by $< \sim 50\%$. Thus, this level of the uncertainty does not affect whether the coefficients are significant or not.
}

\begin{table*}
\caption{Summary of the correlation coefficients calculated for individual cases.} \label{tab-coef}
\begin{threeparttable}
\begin{tabular}{*{4}{l}}
\hline\hline
Case & RX J1713\footnotemark[*] & IC\footnotemark[\dag] & Correlation coefficient\footnotemark[\ddag] \\
\hline
$t_{\rm p}$ ($t_{\rm s}$) and $E_{\rm cut}$  & Y & Y & $ -0.32 \pm 0.07 \pm 0.27  \,(-0.30 \pm 0.07 \pm 0.27) $  \\
  & N & Y & $ -0.62 \pm 0.09 \pm 0.20  \,(-0.26 \pm 0.07 \pm 0.29) $  \\
  & N & N & $ -0.58 \pm 0.10 \pm 0.26  \,(-0.14 \pm 0.10 \pm 0.36) $  \\
 
$t_{\rm p}$ ($t_{\rm s}$) and $E_{\rm br}$  & Y & Y & $ -0.23 \pm 0.08 \pm 0.32  \,(-0.59 \pm 0.06 \pm 0.25) $  \\
  & N & Y & $ -0.55 \pm 0.10 \pm 0.28  \,(-0.58 \pm 0.07 \pm 0.27) $  \\
  & N & N & $ -0.54 \pm 0.12 \pm 0.37  \,(-0.44 \pm 0.14 \pm 0.40) $  \\
 
$t_{\rm p}$ ($t_{\rm s}$) and $R_{\rm GeV}$  & Y & Y & $ -0.27 \pm 0.05 \pm 0.19  \,(-0.37 \pm 0.04 \pm 0.17) $  \\
  & N & Y & $ -0.42 \pm 0.06 \pm 0.18  \,(-0.34 \pm 0.04 \pm 0.18) $  \\
  & N & N & $ -0.28 \pm 0.07 \pm 0.21  \,(-0.09 \pm 0.05 \pm 0.21) $  \\
 
$t_{\rm p}$ ($t_{\rm s}$) and $\hat{L}$  & Y & Y & $ -0.28 \pm 0.04 \pm 0.19  \,(-0.41 \pm 0.03 \pm 0.17) $  \\
  & N & Y & $ -0.46 \pm 0.04 \pm 0.18  \,(-0.38 \pm 0.04 \pm 0.17) $  \\
  & N & N & $ -0.31 \pm 0.06 \pm 0.21  \,(-0.12 \pm 0.06 \pm 0.21) $  \\
 
$t_{\rm p}$ and $t_{\rm s}$  & Y & Y & $ 0.39 \pm 0.04 \pm 0.18 $ \\
  & N & Y & $ 0.51 \pm 0.04 \pm 0.17 $ \\
  & N & N & $ 0.60 \pm 0.05 \pm 0.18 $ \\
  \hline
\end{tabular}

\begin{tablenotes}
\item[*] Include RX J1713.7-3946 (Y) or not (N).
\item[\dag] {Include or not objects with possible IC emission (either $\Gamma_{\rm cut}$ or $\Gamma_{\rm br}$ is less than 2.0).}
\item[\ddag] Value $\pm$ (statistical error) $\pm$ (standard error of the regression) is presented. The value in the parentheses corresponds to the case with $t_{\rm s}$.
\end{tablenotes}

\end{threeparttable}
\end{table*}

\begin{table*}
\caption{Comparison of the correlation coefficients for the IP and RP SNRs.} \label{tab-coef2}
\begin{threeparttable}
\begin{tabular}{*{5}{l}}
\hline\hline
Case & RX J1713\footnotemark[*] & IC\footnotemark[\dag] & Correlation coefficient (IP SNRs)\footnotemark[\ddag] & Correlation coefficient (RP SNRs)\footnotemark[\ddag] \\
\hline
$t_{\rm p}$ ($t_{\rm s}$) and $E_{\rm cut}$  & Y & Y & $ 0.03 \pm 0.10 \pm 0.45  \,(-0.26 \pm 0.13 \pm 0.43) $  & $ -0.33 \pm 0.12 \pm 0.41  \,(-0.01 \pm 0.12 \pm 0.49) $  \\
  & N & Y & $ -0.86 \pm 0.12 \pm 0.16  \,(-0.15 \pm 0.15 \pm 0.50) $  & $ -0.33 \pm 0.12 \pm 0.41  \,(-0.01 \pm 0.12 \pm 0.49) $  \\
  & N & N & $ -0.06 \pm 0.72 \pm 0.07  \,(0.10 \pm 0.69 \pm 0.91) $  & $ -0.33 \pm 0.12 \pm 0.41  \,(-0.01 \pm 0.13 \pm 0.49) $  \\
 
$t_{\rm p}$ ($t_{\rm s}$) and $E_{\rm br}$  & Y & Y & $ -0.23 \pm 0.12 \pm 0.48  \,(-0.71 \pm 0.09 \pm 0.34) $  & $ 0.19 \pm 0.17 \pm 0.56  \,(0.22 \pm 0.20 \pm 0.56) $  \\
  & N & Y & $ -0.73 \pm 0.13 \pm 0.37  \,(-0.68 \pm 0.10 \pm 0.42) $  & $ 0.19 \pm 0.18 \pm 0.56  \,(0.22 \pm 0.20 \pm 0.56) $  \\
  & N & N & $ -1.00 \pm 0.04 \pm 0.00  \,(-1.00 \pm 0.04 \pm 0.00) $  & $ 0.19 \pm 0.17 \pm 0.56  \,(0.22 \pm 0.20 \pm 0.56) $  \\
 
$t_{\rm p}$ ($t_{\rm s}$) and $R_{\rm GeV}$  & Y & Y & $ -0.04 \pm 0.06 \pm 0.27  \,(-0.23 \pm 0.05 \pm 0.25) $  & $ -0.34 \pm 0.11 \pm 0.31  \,(0.14 \pm 0.11 \pm 0.33) $  \\
  & N & Y & $ -0.24 \pm 0.07 \pm 0.27  \,(-0.17 \pm 0.05 \pm 0.26) $  & $ -0.34 \pm 0.11 \pm 0.31  \,(0.14 \pm 0.11 \pm 0.33) $  \\
  & N & N & $ -0.10 \pm 0.08 \pm 0.33  \,(-0.05 \pm 0.07 \pm 0.33) $  & $ -0.34 \pm 0.11 \pm 0.31  \,(0.14 \pm 0.11 \pm 0.33) $  \\
 
$t_{\rm p}$ ($t_{\rm s}$) and $\hat{L}$  & Y & Y & $ -0.05 \pm 0.05 \pm 0.27  \,(-0.20 \pm 0.05 \pm 0.25) $  & $ -0.43 \pm 0.14 \pm 0.29  \,(-0.03 \pm 0.14 \pm 0.33) $  \\
  & N & Y & $ -0.28 \pm 0.06 \pm 0.27  \,(-0.13 \pm 0.05 \pm 0.27) $  & $ -0.43 \pm 0.13 \pm 0.29  \,(-0.03 \pm 0.14 \pm 0.33) $  \\
  & N & N & $ -0.13 \pm 0.07 \pm 0.33  \,(-0.00 \pm 0.07 \pm 0.33) $  & $ -0.43 \pm 0.13 \pm 0.29  \,(-0.02 \pm 0.14 \pm 0.33) $  \\
 
$t_{\rm p}$ and $t_{\rm s}$  & Y & Y & $ 0.22 \pm 0.05 \pm 0.26 $ & $ 0.68 \pm 0.05 \pm 0.24 $ \\
  & N & Y & $ 0.41 \pm 0.06 \pm 0.25 $ & $ 0.68 \pm 0.05 \pm 0.24 $ \\
  & N & N & $ 0.54 \pm 0.06 \pm 0.28 $ & $ 0.68 \pm 0.05 \pm 0.24 $ \\
  \hline
\end{tabular}

\begin{tablenotes}
\item[*] Include RX J1713.7-3946 (Y) or not (N).
\item[\dag] {Include or not objects with possible IC emission (either $\Gamma_{\rm cut}$ or $\Gamma_{\rm br}$ is less than 2.0).}
\item[\ddag] Value $\pm$ (statistical error) $\pm$ (standard error of the regression) is presented. The value in the parentheses corresponds to the case with $t_{\rm s}$.
\end{tablenotes}

\end{threeparttable}
\end{table*}

\section{Discussion}\label{sec-discussion}

\subsection{Comparison of $E_{\rm cut}$ and $E_{\rm br}$ between the measurements and theoretical calculations}\label{subsec-esccut}
In Figure \ref{fig-di}, the SNRs with RPs show relatively larger photon indices. Considering the larger $t_{\rm p}$ or $t_{\rm s}$ of the objects with RPs, the older SNRs may generally have softer gamma-ray spectra.
This is consistent with the result in \citet{acero16}, which shows a relation between the SNR age and the 1--100 GeV photon index. This implies that the spectral softening due to particle escape also affects the spectral index in addition to $E_{\rm cut}$ or $E_{\rm br}$.
Since $t_{\rm p}$ and $t_{\rm s}$ are calculated independently, the consistency between them within a factor for most of the objects (Figure \ref{fig-ps}; Section \ref{sec-sysanalysis}) suggests that $t_{\rm p}$ and $t_{\rm s}$ roughly represent the same value, which should be the SNR age.
The difference between $t_{\rm p}$ and $t_{\rm s}$ is most simply explained as due to a wrong estimation of the post-shock density from X-ray observations.
This is not irrational since an estimation of $n_{\rm e}$ in an X-ray analysis is always accompanied by a potential uncertainty of the filling factor of the X-ray emitting plasma, which we ignore in this work.
If the post-shock density is overestimated by $\sim$3 times, the difference of the factor of $\sim$6 is explained. Thus, this may indicate that X-ray emitting plasmas have $\sim$3 times larger densities than the average values in post-shock regions.
Hereafter, we assume that $t_{\rm p}$ and $t_{\rm s}$ represent the SNR age within a factor.

In Figure \ref{fig-pc} and \ref{fig-pb}, {we see only a few data points are close to the theoretical calculations in a Bohm diffusion case (with $B_0 = 5~\mu$G and $n = 0.4$ cm$^{-3}$ in the analytical, and $B_0 = 4~\mu$G and $n = 0.4$ cm$^{-3}$ in the numerical calculations). If we assume that the theoretical calculations in a Bohm diffusion case explain the average trends of the data (gray regions), unrealistic values of $B_0 < 0.1 ~\mu$G or $n > 10^4$ cm$^{-3}$ will be required \citep{ptuskin03}. Thus it is hard to expect that an assumption of a Bohm diffusion can generally be applied to these SNRs.} 
The data trends seen in Figure \ref{fig-pc} and \ref{fig-pb} show that the data can be reproduced with the theoretical predictions for acceleration conditions in a thin interstellar medium (``nonlinear damping (analytical)'' and ``nonlinear damping (numerical)'' cases in Figure \ref{fig-pc} and \ref{fig-pb}), or under ion-neutral collision (``ion-neutral damping (analytical)'' in Figure \ref{fig-pc} and \ref{fig-pb}).
The time evolution of $E_{\rm br}$ is generally lower than that of $E_{\rm cut}$, so that it prefers the ``ion-neutral damping (analytical)'' case. Since the spectral shape at the maximum energy of the accelerated particles depends on whether escaping particles emit gamma-rays as well, we can not determine which of $E_{\rm cut}$ or $E_{\rm br}$ is proper to describe the escape environments.
Thus, it is difficult to conclude which escape condition best reproduces the observational trends.
The significant dispersions of $E_{\rm cut}$ or $E_{\rm br}$ of 2--3 orders of magnitude even at the same $t_{\rm p}$ or $t_{\rm s}$ values might be due to wrong estimations of $t_{\rm p}$ and $t_{\rm s}$ (or possibly $E_{\rm cut}$ and $E_{\rm br}$).
However, these might be due to an intrinsic variety of escape environments: ambient density and ionization state, and $B_0$.
Therefore, we suggest a possibility that, the complicated spectral shape of the Galactic cosmic rays observed on Earth (e.g., \cite{lipari19}) is due to this intrinsic variety of the escape environments,
since even if the spectral indices of accelerated particles are the same, those of escaping particles differ significantly depending on the escape timescale \citep{ohira10, ohira11b}.

\subsection{Escape timescale estimated with $\hat{L}$}
Figure \ref{fig-ph} and \ref{fig-psl} exhibit negative correlations between $t_{\rm p}$ or $t_{\rm s}$ and $R_{\rm GeV}$ or $\hat{L}$ for most cases of the sample selection (Table \ref{tab-coef}).
The theoretical predictions shown with magenta lines \citep{yasuda19} are roughly consistent with the data trends, although these models can not explain the time evolutions of $E_{\rm cut}$ and $E_{\rm br}$ (Figure \ref{fig-pc} and \ref{fig-pb}).
This is probably because $\hat{L}$ is insensitive to the spectral shape around the maximum energy of the accelerated particles, as we discussed in Section \ref{sec-sysanalysis}.

From Figure \ref{fig-psl} (a) and (b), we can estimate the timescale of the particle escape, in which $\hat{L}$ becomes $\sim$1/3 for example, to be two orders of magnitude increase of time.
{Under an assumption that gamma-ray luminosities take maximum at around 1 kyr \citep{yasuda19}, the decrease timescale of total confined proton energy is roughly estimated as $\sim$100 kyr.}
With a fixed gamma-ray photon index of 2.0, this timescale corresponds to the decrease of $E_{\rm cut}$ to $\sim$10 GeV, which is consistent with Figure \ref{fig-pc} and \ref{fig-pb}.
This decreasing trend of the luminosity can not be seen in Figure \ref{fig-pl} (c) or (d). This should be due to an intrinsic variety of the explosion kinetic energies and/or the surrounding environments, which can not be excluded only by dividing the luminosities by the ambient densities.
We note that, the narrower distributions of the luminosities divided by the ambient densities compared to the luminosity distribution may in fact indicate that the intrinsic variety is somewhat reduced in Figure \ref{fig-pl} (c) and (d).

\subsection{Implication on the origins of RPs}
In Section \ref{subsec-esccut}, we suggest that $t_{\rm p}$ represents the true SNR age within a factor.
Since RP ages do not necessarily correspond to the SNR ages themselves considering their unclear origins, it is surprising that Figure \ref{fig-ps} shows no clear separation between the distributions of the IP and the RP SNRs ($< 2~\sigma$ significance in terms of the difference of the factor $f$; see Section \ref{sec-sysanalysis}).
Therefore, this may provide us with an important clue for the origins of RPs, which suggests that very early stages of SNR evolution are responsible for generation of RPs. \citep{itoh89, katsuragawa19, zhang19}.

\subsection{An outlier: RX J1713.7-3946}\label{subsec-1713}
RX J1713.7-3946 is an outlier in terms of $t_{\rm p}$, since it clearly has a too large value of $\sim$160 kyr. This may be because of its unnaturally high ionization timescale of $\sim 5 \times 10^{11}$ cm$^{-3}$ s$^{-1}$ considering its kinematic age of $\sim$1600 year \citep{katsuda15}. Alternatively, this may be due to a systematic uncertainty of its post-shock density, which is hard to estimate because of the very low flux of the thermal component buried in the bright non-thermal emission. Thus we suggest two possibilities: one is that it has undergone a rapid ionization in a dense environment in its early stages and then somehow the thermal plasma was rarefied to become thin. The other is that the current thermal plasma density is actually much higher, but is underestimated because of the overestimation of its volume.
The correlation coefficients without RX J1713.7-3946 are also presented in Table \ref{tab-coef}, most of which indicate significant correlations, although the values with and without RX J1713.7-3946 are consistent within uncertainties.

Most of the coefficients calculated without the objects with possible IC gamma-rays are significant for the plots with $t_{\rm p}$, which indicates an improvement of the correlations (Table \ref{tab-coef}).
However, considering the fact that most of the cases without RX J1713.7-3946 already show significant coefficients (Table \ref{tab-coef}), these improvements are attributed to the lack of RX J1713.7-3946.
{Moreover, in the cases with $t_{\rm s}$, almost no significant correlations are found if the possible IC objects are excluded.}
Thus, the contribution of the possible IC objects can not be distinguished in this work.

\section{Conclusion}\label{sec-summary}

We conducted a systematic analysis on the gamma-ray emitting SNRs using their gamma-ray emission and thermal plasma properties, in order to estimate the timescales of high-energy particle escape.
The two ages $t_{\rm p}$ and $t_{\rm s}$ are found to be close within $\sim$ a factor, suggesting that they roughly represent the same value, which should be the SNR age.
The $E_{\rm cut}$ and $E_{\rm br}$ are found to decrease with increasing $t_{\rm p}$ {in several cases of the sample selection.} The comparisons with several analytical/numerical calculations suggest that, although {acceleration with a Bohm diffusion seems to be hard to explain most of the sample}, that under a shock-ISM or a shock-cloud interaction can reproduce the data trends.
Also, by comparing the normalized gamma-ray luminosity $\hat{L}$ to $t_{\rm p}$ and $t_{\rm s}$, we estimate the decrease timescale of the total energy of confined protons as $\sim$100 kyr.
We suggest that $t_{\rm p}$ represents the true SNR age within a factor and that very early stages of SNR evolution are responsible for generation of RPs. This can be crucial to understanding the origins of RPs.

\begin{ack}

We are grateful to the anonymous referee for providing us with valuable advice, especially on the statistical treatment of the data.
We deeply thank R. Brose and H. Yasuda for providing us with their simulation results and for discussions on gamma-ray spectral shape of escaping particles.
We also thank S. Katsuda for the information on the thermal plasma in RX J1713.7-3946.
HS is supported by JSPS Research Fellowship for Young Scientist (No. 19J11069).
This research was partially supported by JSPS KAKENHI Grant Nos. 19K03908 (AB), 18H01232 (RY), JP16K17702 and JP19H01893 (YO),
the Grant-in-Aid for Scientific Research on Innovative Areas “Toward new frontiers: Encounter and synergy of state-of-the-art astronomical detectors and exotic quantum beams” (18H05459; AB)
and Shiseido Female Researcher Science Grant (AB).
YO is supported by Leading Initiative for Excellent Young Researchers, MEXT, Japan.
RY deeply appreciates Aoyama Gakuin University Research Institute for helping our research by the fund.

\end{ack}

\end{document}